\documentclass[smallextended, final]{svjour3}

\makeatletter
\def\makeheadbox{\relax}
\makeatother

\smartqed  
\usepackage{graphicx}
\usepackage{natbib}
\usepackage{url}
\usepackage{multirow}
\usepackage{array,booktabs}
\usepackage{tcolorbox}
\usepackage{makecell}
\usepackage{tabularx}
\definecolor{cellh}{gray}{1.0} 
\usepackage[colorlinks, linkcolor=blue, citecolor=blue, urlcolor=blue]{hyperref}

\let\cite\citep

\usepackage{pifont}   
\newcommand{\cmark}{\ding{51}} 
\newcommand{\xmark}{\ding{55}} 

\begin{document}

\title{Where are the Hidden Gems? Applying Transformer Models for Design Discussion Detection}


\author{Lawrence Arkoh \and Daniel Feitosa \and Wesley K. G. Assunção
}


\institute{
Lawrence Arkoh \at
North Carolina State University, Raleigh, USA.
\email{larkoh@ncsu.edu}
\and
Daniel Feitosa \at 
University of Groningen, Groningen, Netherlands. 
\email{d.feitosa@rug.nl}  
\and
Wesley K. G. Assunção \at
North Carolina State University, Raleigh, USA.
\email{wguezas@ncsu.edu}  
}

\date{Received: date / Accepted: date}

\maketitle

\begin{abstract}
Design decisions are at the core of software engineering and appear in Q\&A forums, mailing lists, pull requests, issue trackers, and commit messages. Design discussions spanning a project’s history provide valuable information for informed decision-making, such as refactoring and software modernization. Machine learning techniques have been used to detect design decisions in natural language discussions; however, their effectiveness is limited by the scarcity of labeled data and the high cost of annotation.
Prior work adopted cross-domain strategies with traditional classifiers, training on one domain and testing on another. Despite their success, transformer-based models, which often outperform traditional methods, remain largely unexplored in this setting.
The goal of this work is to investigate the performance of transformer-based models (i.e., BERT, RoBERTa, XLNet, LaMini-Flan-T5-77M, and ChatGPT-4o-mini) for detecting design-related discussions. To this end, we conduct a conceptual replication of prior cross-domain studies while extending them with modern transformer architectures and addressing methodological issues in earlier work. The models were fine-tuned on Stack Overflow and evaluated on GitHub artifacts (i.e., pull requests, issues, and commits). BERT and RoBERTa show strong recall across domains, while XLNet achieves higher precision but lower recall. ChatGPT-4o-mini yields the highest recall and competitive overall performance, whereas LaMini-Flan-T5-77M provides a lightweight alternative with stronger precision but less balanced performance.
We also evaluated similar-word injection for data augmentation, but unlike prior findings, it did not yield meaningful improvements. Overall, these results highlight both the opportunities and trade-offs of using modern language models for detecting design discussion.

\end{abstract}

\section{Introduction}
\label{intro}

The foundation of a software system is its architectural design and the decisions made to arrive at its current state~\cite{wan2023software,bucaioni2024continuous}.\footnote{In this paper, the term ``design'' refers explicitly to software structural design, and design decisions are architecture-related text, as described by~\citet{brunet2014developers}.} Thus, during the software's development life cycle, developers need to make many design decisions~\cite{alkadhi2018developers}. These decisions are usually related to functional requirements, non-functional requirements, or technological constraints, which, in turn, affect the architecture~\cite{ameller2012non}. When made incorrectly or suboptimally, software design decisions can lead to software degradation and incur financial and maintenance costs, namely technical debt~\cite{li2015systematic, da2017using, skryseth2023technical}, pushing software systems deep into the legacy system zone~\cite{assunccaocontemporary}.

To maintain and evolve aging systems, ensuring their value over time, developers should be able to recover and understand the design decisions made during software development~\cite{josephs2022towards}. 
Modern software development platforms\footnote{\url{https://github.com}, \url{https://gitlab.com}, and \url{https://www.atlassian.com/software/jira}} offer features such as pull requests and issues management, allowing developers to review code, raise issues with a given piece of code, and then have discussions to resolve them~\cite{yang2022developers}. Additionally, software design decisions can also be discussed in mailing lists \cite{viviani2019locating} and commit messages, where developers document their intentions behind code changes~\cite{tian2022makes,dhaouadi2024rationale}.
Furthermore, developer forums such as Stack Overflow\footnote{\url{https://Stack Overflow.com}} are yet another platform where developers exchange information and discuss architecture decisions (among other topics)~\cite{de2023characterizing}.

Pull requests, issues, emails, messages, Q\&A forums, or any discussion medium, are valuable sources of information related to design discussion~\cite{oliveira2024understanding,viviani2019locating,brunet2014developers}. 
These pieces of information enable an understanding of software architectures and the history of their design evolution. Consequently, this information supports system maintenance and evolution (i.e., code refactoring, redesign, migration, or modernization) to meet new demands. 
However, exploiting multiple communication channels has been a challenge due to a lack of tools and processes for identifying such discussions~\cite{maarleveld2023maestro, mehrpour2023survey}. 
Because these discussions are seldom formally documented, studies have shown that the rationale of decisions related to a software's architecture may vaporize with time~\cite{borrego2019towards,li2022understanding}. Also, unfortunately, current tools and practices poorly track which discussions are design decision-related.



A few studies have proposed approaches based on machine learning (ML) for classifying textual discussions from different sources as design-related or non-design-related discussions~\cite{brunet2014developers, viviani2019locating, shakiba2016fourd, maarleveld2023maestro}. These studies trained ML classifiers and evaluated them on datasets from the same source. That is, both the training and testing datasets were derived solely from pull requests, issues, or commit messages. 
The problem with making these approaches applicable in practice is that, in addition to their sensitivity to the source dataset type, there is limited availability of labelled datasets, which further hinders the performance and generalizability of the tools.

To address the problem mentioned above, \citet{mahadi2022conclusion} curated a dataset from Stack Overflow discussions and trained traditional classifiers on identifying similar design discussions in commit messages, pull requests, and issue discussions. This approach is referred to as \emph{cross-domain dataset classification}, which involves training the model on a domain (in this case, a Stack Overflow discussion) and applying it to classify data from other domains (e.g., commits, pull requests, issues)~\cite{zimmermann2009cross}. The authors also employed \emph{similar-word injection}, a data augmentation strategy that enhances classifier performance by reinforcing its capacity to generalize and transfer contextual information from the Stack Overflow dataset to commit messages, pull requests, and issue discussions.
\citet{mahadi2022conclusion} highlighted challenges in the form of low ML model performance when trained on data from one domain (i.e., Stack Overflow) and tested on data from different domains, despite both containing design discussions in natural language text.

To advance the state of the art in design discussion detection and pave the way for building better tools, this present work aims to \textit{evaluate transformer-based models for design discussion identification}. 
Motivated by the strong performance of transformer-based language models (e.g., BERT~\cite{bert} and ChatGPT~\cite{chatgpt}) in understanding and classifying natural language, we addressed the limitations and extended the work by \citet{mahadi2022conclusion} by conducting a \textit{conceptual replication}~\cite{cruz2019replication, gomez2014understanding}. Specifically, we explored whether five state-of-the-art transformer-based models, namely BERT~\cite{bert}, RoBERTa~\cite{liu2019roberta}, and XLNet~\cite{yang2019xlnet} (all discriminative models), as well as LaMini-Flan-T5-77M~\cite{lepagnol2024smalllanguagemodelsgood} and ChatGPT-4o-mini~\cite{chatgpt} (generative language models), can effectively generalize their ability to distinguish between design-related and non-design discussions in a cross-domain setting. To this end, we fine-tuned the models on Stack Overflow discussions and tested them on similar categories of text drawn from GitHub pull requests~\cite{viviani2019locating,brunet2014developers}, code comments~\cite{da2017using}, issues and commit messages~\cite{brunet2014developers}. Our key findings are:

\begin{itemize}
    \item Transformer-based models can accurately detect design-related discussions when training and testing data originate from the same domain. 
    
    \item A transformer-based model with a reduced dataset, namely Stack Overflow questions excluding answers and comments, can be sufficient for robust detection while reducing fine-tuning time and computational costs.

    \item Despite outperforming traditional ML classifiers, transformers exhibit performance degradation, especially in balanced precision and recall, when applied across domains (e.g., Stack Overflow to GitHub). This highlights the need for careful use of domain adaptation techniques.

    \item ChatGPT-4o-mini consistently achieved higher recall across datasets, making it effective for exploratory analyses where capturing design discussions is critical, while LaMini-Flan-T5-77M offered stronger precision and efficiency, positioning it as a practical choice for resource-constrained cases.

    \item Tool builders in this domain should consider the trade-off between recall and precision, as no single model is universally optimal across settings.
\end{itemize}

Given these findings, the contributions of our research provide better support for software engineering practitioners by enabling a lightweight, automated means of detecting design-related discussions. This improves design traceability and team alignment without requiring manual effort, as these models can identify design decisions without additional tedious tasks for humans. The detected software design-related discussions can enable practitioners to enhance documentation, streamline onboarding, and support knowledge transfer, ultimately guiding maintenance and modernization efforts. By making design decisions more accessible, our work contributes to a deeper understanding of a given software's architecture and aids in its modernization over time.

In the remainder of this paper, Section~\ref{sec:related-work} presents the related work, followed by our study design in Section~\ref{sec:methodology}. Section~\ref{sec:results} presents the results, and Section~\ref{sec:discussion} presents the discussion and implications, with the threats to validity. We conclude the paper in Section~\ref{sec:conclusion}.

\section{Related Work}
\label{sec:related-work}

Despite its importance, the detection and classification of software design decisions remain relatively underexplored. Existing studies report a lack of systematic documentation and automated techniques, limited empirical evidence, and immature classification approaches in this area~\cite{assunccaocontemporary, mondal2022survey, bhat2017automatic, bi2021mining, su2026evaluating}. 

\citet{wijerathna2022mining} proposed a hybrid approach to mine design patterns and contextual data by combining unsupervised and supervised ML techniques. It employs unsupervised methods to generate vector representations of design patterns, capturing their semantic features. Then, a Support Vector Classifier (SVC) is trained on a manually labeled Stack Overflow dataset to categorize posts into different design patterns. The SVC achieved an ROC-AUC\footnote{Area Under the Receiver Operating Characteristic Curve measures the classifier's ability to distinguish between different class labels, with performance values ranging from 0 (worst), through 0.5 (eq. to random guessing),  to 1 (best)~\cite{sofaer2019area}.} of 0.87, with a precision of 0.815 and a recall of 0.880. Further incorporation of collocations into the test data led to a slight improvement in performance, with precision and recall increasing to 0.815 and 0.882, respectively. \citet{zhao2024drminer} studied prompt tuning with large language models to mine latent design rationales from Jira issue logs. This method yielded a 22\% improvement in F1 over baseline techniques in their study.

\citet{brunet2014developers} examined design-related discussions across GitHub commits, issues, and pull requests. They manually labeled 1,000 samples as related or unrelated to structural design, then trained a Decision Tree classifier using this data. The model achieved an accuracy of 94 ± 1\% and was later applied to over 102,000 discussion samples, revealing that approximately 25\% were related to software structural design.

\citet{da2017using} developed a method to automatically detect comments related to design requirements and technical debt. They used a Maximum Cross Entropy classifier trained on labeled code comments, which achieved an average F1 of 0.62 on their test set.
\citet{viviani2019locating} manually labeled paragraphs from GitHub pull requests from open-source projects as design-related or not. Then, they trained multiple classifiers, with the best-performing model (Random Forest) achieving ROC-AUC scores of 0.87 on the training set and 0.81 on a cross-project test set.

All the studies above focus on detecting design-related discussions within a single domain by training and testing models on data from the same source. In contrast, \citet{mahadi2022conclusion} investigated the cross-domain performance of several ML classifiers, including Linear SVM, Logistic Regression, and RBF SVM. They trained models on Stack Overflow data and evaluated them on datasets from other domains, such as code comments, commit messages, and pull request discussions. 
The classifiers performed well on a similar Stack Overflow dataset, achieving an ROC-AUC of 0.882, which demonstrates high performance in differentiating design-related texts from non-design-related texts. However, the performance of the models deteriorated significantly on cross-domain datasets. 
Specifically, when tested on data from \citet{brunet2014developers}, the ROC-AUC dropped to 0.632, and even lower scores were observed on the datasets from \citet{da2017using}, \citet{viviani2019locating}, and \citet{shakiba2016fourd}, with ROC-AUC scores of 0.496, 0.483, and 0.513, respectively. These scores indicate that the models performed poorly in cross-domain classification, achieving only random guessing accuracy. To address this, they applied similar word injection data augmentation to the dataset from \citet{brunet2014developers}, resulting in a ROC-AUC score of 0.7985, which represents a 26\% improvement. 

While a previous study employed transformer models~\cite{zhao2024drminer}, they did not address the challenge of cross-domain dataset classification. The few ones that explored cross-domain design discussion primarily relied on traditional NLP techniques and classical ML algorithms~\cite{da2017using,viviani2019locating,mahadi2022conclusion}, which generally underperform compared to modern transformer-based models.

\section{Study Design}
\label{sec:methodology}


The goal of our study is to investigate the stability of transformer-based models in identifying design discussions across datasets from different domains. We conduct a conceptual replication of the study by~\citet{mahadi2022conclusion}, described in the previous section, by evaluating the performance of transformer-based models on multiple independently collected datasets. Specifically, we assess model performance using datasets provided by~\citet{brunet2014developers}, \citet{viviani2019locating}, and~\citet{da2017using}. In doing so, we also address methodological issues in the original study, as discussed in Section~\ref{sec:evaluation}.
We structure our study around the following two research questions (RQs):

\vspace{2mm}\noindent
\textbf{RQ1. How effective are transformer-based models at identifying software design-related discussions in cross-domain developer discussions?}
This RQ assesses the capability of modern transformer-based models (presented in Section~\ref{sec:transformers}) in identifying software design-related discussions from developer artifacts such as commit messages, pull requests, and issue discussions. By answering this RQ, we aim to understand whether these models can serve as reliable tools for automatically extracting design knowledge, potentially improving documentation and traceability in software projects. The goal is to understand which models perform best and thereby offer insights for both researchers and practitioners looking to adopt NLP techniques in design discussion detection.

\vspace{2mm}\noindent
\textbf{RQ2. Does similar-word injection enhance the cross-domain performance of transformer-based models in detecting software design-related discussions?}  This RQ investigates whether applying \textit{similar-word injection}~\cite{long2024data} augmentation strategy (i.e., replacing words in unseen inputs with semantically related terms), can improve the cross-domain generalization of transformer-based models in detecting software design-related discussions. Generalization remains a significant challenge in software engineering NLP tasks, as terminology, phrasing, and contextual structure often vary across projects. By enriching test inputs with lexical variations, this approach aims to make models more robust to domain shift without requiring repetitive model retraining~\cite{lu2022improved,shanmugam2025test}. In this RQ, we assess the impact of augmentation on classification accuracy in cross-domain settings and explore whether it offers a practical alternative for improving performance.
\vspace{2mm}

In the following subsections, we detail the methodology of our replication study. 
To enable further research and replication, the data and code used in our study is publicly available as supplementary material~\cite{supplementary}.

\subsection{Data Collection and Preprocessing}

Given the time and effort required to label data for text classification tasks~\cite{fredriksson2020data}, we reused datasets from prior studies. 
Table~\ref{tab:datasets} presents a summary of the datasets gathered from four studies, which we used in our replication.
To obtain a large labeled dataset, \citet{mahadi2022conclusion} relied on Stack Overflow. In their study, texts from questions, answers, and comments were labeled as \textit{design} or \textit{general} based on the user-supplied tags added when the questions were asked. \citet{brunet2014developers} collected text from commit messages, issues, and pull requests from GitHub and classified them as \textit{design} or \textit{non-design} related discussions. Similarly, \citet{viviani2019locating} and \citet{da2017using} also labeled text from pull requests and commit messages as \textit{design} or \textit{non-design} related discussions. 


\begin{table}[!tp]
\centering
\caption{Datasets of design dicussions used in our study.}
\addtolength{\tabcolsep}{-2pt}
\begin{tabular}{l|p{2.7cm}|r|r|r }
\toprule
\multirow{2}{*}{\textbf{Reference}} & \multirow{2}{*}{\textbf{Source(s)}} & \multicolumn{2}{c|}{\textbf{Design related?}}& \multirow{2}{*}{\textbf{Total}}\\ \cmidrule{3-4}
& & \textbf{Yes} & \textbf{No} & \\ 
\midrule
Mahadi~\cite{mahadi2022conclusion} & Stack Overflow questions, comments, answers & 115,000 & 115,000 & 230,000 \\ \midrule
Brunet~\cite{brunet2014developers} & GitHub commit messages, issues, and pull requests  & 246 & 754 & 1,000 \\ \midrule
Viviani~\cite{viviani2019locating} & GitHub pull requests& 6,250 & 2,365 & 8,615 \\ \midrule
SATD~\cite{da2017using} & Code comments & 2,599 & 1,183 & 3,782 \\ 
\bottomrule
\end{tabular}
\label{tab:datasets}
\end{table}

We followed the same approach of existing studies to pre-process the data by removing some special characters and other data-specific content (e.g., thread-start, thread-end, FIXME, TODO, web URLs) that would not have any impact on the fine-tuning~\cite{skryseth2023technical,mahadi2022conclusion}.
The datasets provided by \citet{mahadi2022conclusion} and \citet{brunet2014developers} were already cleaned, not requiring further action. However, we performed these cleaning steps for the data provided by \citet{da2017using} and \citet{viviani2019locating}. We note that using transformer-based models does not require additional steps like removal of stop words and parts-of-speech tagging due to the nature of their training. In fact, further pre-processing could impact their performance~\cite{fan2023stop}.

\subsection{Model Selection and Fine-Tuning}
\label{sec:transformers}

The introduction of Bidirectional Encoder Representations from Transformers (BERT)~\cite{bert} established a new standard for understanding natural language. Their bi-directional attention mechanism enables it to capture semantic meaning and sentence structure effectively~\cite{zhao2024drminer}. BERT's widespread success in various classification tasks across domains~\cite{pan2021automating, shang2024analyzing, ajagbe2022retraining, li2024fine, skryseth2023technical, josephs2022towards} highlights its robust generalization and accuracy. Motivated by these strengths, we apply transformer-based models to the task of cross-domain software design discussion detection. Although models like RoBERTa~\cite{liu2019roberta} and XLNet~\cite{yang2019xlnet} have demonstrated superior performance in general classification benchmarks~\cite{izadi2022predicting, nadeem2021automatic, arabadzhieva2022comparison, kumar2021fake}, their effectiveness in the specific context of identifying design-related communication within software repositories remains underexplored. In our study, we used BERT (\textit{bert-base-uncased})~\cite{bert}, RoBERTa (\textit{roberta-base})~\cite{liu2019roberta}, and XLNet (\textit{xlnet-base-cased})~\cite{yang2019xlnet} through the HuggingFace library.\footnote{\url{https://huggingface.co}} 

We also include ChatGPT-4o-mini~\cite{chatgpt} in our study, which is a large-scale generative transformer model that has shown strong performance in software engineering tasks~\cite{widyasari2025explaining, sun2023automatic}. Its conversational fine-tuning allows it to capture nuanced design-related discussions with high recall and competitive overall effectiveness.
In addition to large-scale transformer models, we also consider LaMini-Flan-T5-77M~\cite{lamini-lm}, a smaller and resource-efficient model, which has demonstrated competitive results under limited-resource settings~\cite{lepagnol2024smalllanguagemodelsgood} as opposed to the high computational costs of large language models~\cite{patel2024characterizing, shekhar2024towards, liagkou2024cost}.

We applied transfer learning and fine-tuning to adapt each model to our datasets. For ChatGPT-4o-mini and LaMini-Flan-T5-77M, which require natural language prompting, we employed the prompt and context template shown below. The prompt was curated to align with the design discussion classification criteria introduced by~\citet{brunet2014developers}, which, to the best of our knowledge, represents the earliest study to formulate the detection of software design discussions as a supervised machine learning classification task. This conceptual definition has since been adopted by subsequent work, including the replicated study by~\citet{mahadi2022conclusion}. Grounding our prompt in this established formulation ensures methodological consistency with prior studies and enables fair comparison and replication rather than relying on ad-hoc or task-specific phrasing.
\begin{tcolorbox}
    Classify the following text as \textbf{design} if it is related to software structural design or \textbf{general} otherwise: \texttt{\{\{TEXT TO CLASSIFY\}\}}
\end{tcolorbox}

In contrast, models such as BERT, RoBERTa, and XLNet were fine-tuned using structured input representations without the need for natural language prompts.

As part of a conceptual replication of the study by \citet{mahadi2022conclusion}, we reused their Stack Overflow training dataset to fine-tune all models, enabling direct comparison with prior results while isolating the impact of transformer-based architectures. Originally, their dataset consisted of 200,000  samples for training, 30,000 samples for testing, and 30,000 samples for validation. However, we observed that the test and validation samples were duplicates, which is a methodological issue, so we discarded the validation samples and only considered the remaining 230,000 for our study. 
We selected the same 200,000 used for fine-tuning and validation (80\% and 20\% split) as well as the same test samples for model testing, which were never exposed to our model during fine-tuning. The full decomposition of the Stack Overflow dataset is presented in Table~\ref{tab:primary_dataset}.

\begin{table}[!tp]
\centering
\caption{Summary of Primary Dataset (Stack Overflow Training and Validation only).}
\begin{tabular}{l|r|r|r}
\toprule
\textbf{Category} & \textbf{Design} & \textbf{Non-Design} & \textbf{Total} \\ 
\midrule
Questions & 50,000 & 50,000 & 100,000 \\ 
Answers & 20,000 & 20,000 & 40,000 \\
Comments & 30,000 & 30,000 & 60,000 \\ 
Combined & 100,000  & 100,000  & 200,000  \\ 
\bottomrule
\end{tabular}
\label{tab:primary_dataset}
\end{table}

\begin{figure}[!tp]
    \centering    
    \includegraphics[width=0.90\textwidth]{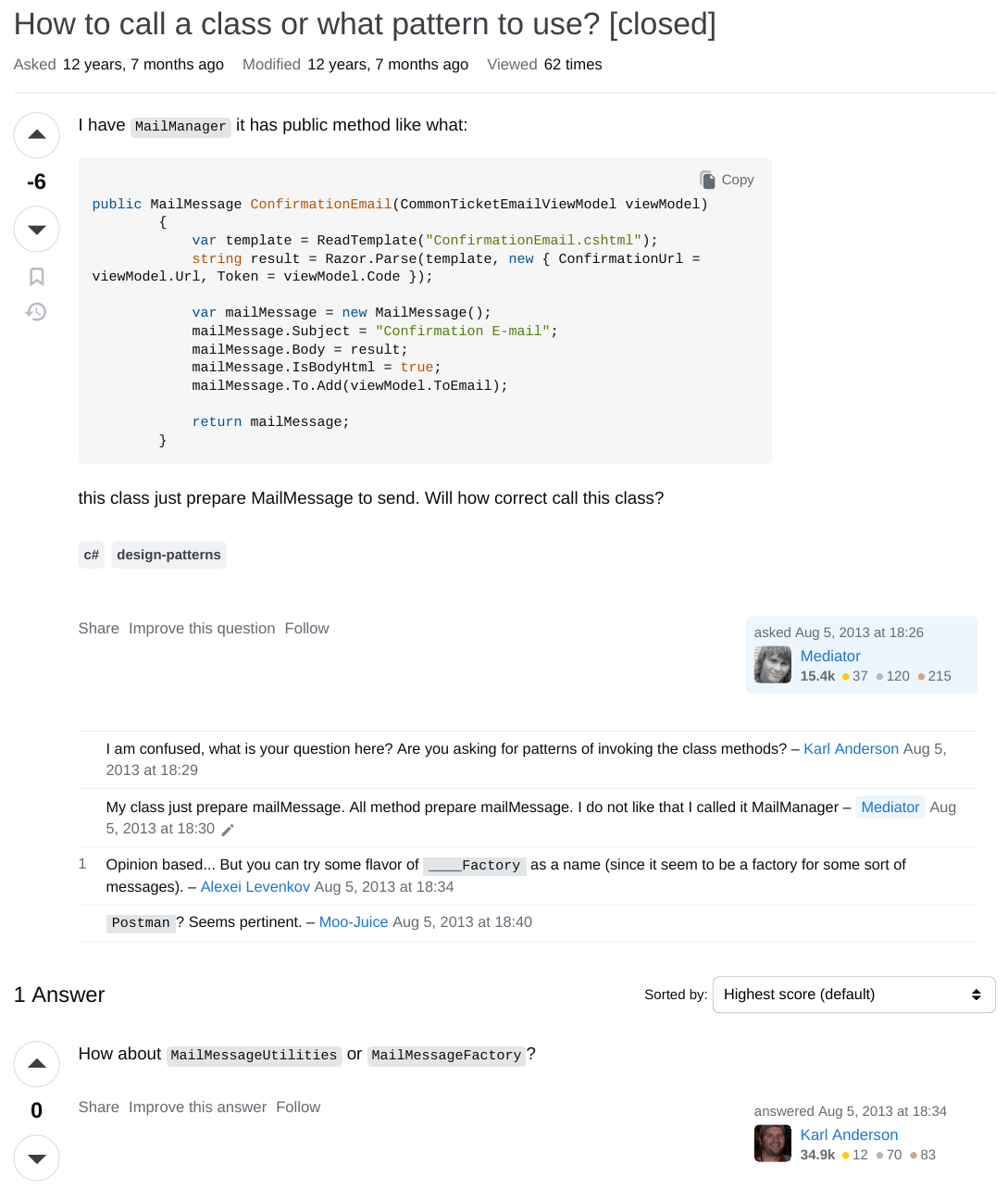}
    \caption{Sample Stack Overflow question about design, but with non-design related comments and answer}
    \label{fig:sample_stack_overflow}
\end{figure}

Furthermore, \citet{mahadi2022conclusion} labeled their data by identifying questions tagged with software architecture or design decision-related terms, categorizing them as ``Design'', while those without such tags were labeled as ``General''. However, based on our observations of how users interact on Stack Overflow, a question tagged as design-related does not necessarily imply that its associated answers and comments are related to design. Additionally, tagging on Stack Overflow is applied only to parent questions, not to their corresponding answers or comments. An example of this is illustrated in Figure~\ref{fig:sample_stack_overflow}, where the question description pertains to a design-related topic, but some of the comments and answers lack any discussion or context related to design patterns.\footnote{\url{https://Stack Overflow.com/questions/18065115/how-to-call-a-class-or-what-pattern-to-use}}
Based on that, we conducted and additional experiment. First, we fine-tuned our models on the combined questions, answers, and comments dataset. Second, we fine-tuned the models on the dataset, which comprises questions only, and compared the two results to verify the following hypothesis: 

\begin{tcolorbox}
\textit{Discarding answers and comments alongside questions in the fine-tuning data does not significantly affect the performance of models in detecting design discussions.}
\end{tcolorbox}

\subsection{Experimentation Setup}
\label{sec:evaluation}

We used a separate, balanced subset of the Stack Overflow dataset comprising 30,000 textual discussions (15,000 design-related and 15,000 non-design-related) drawn from the full 230,000-text corpus for the evaluation on the same domain (i.e., Within Dataset). Then, the other datasets (Brunet, Viviani, and SATD) were used to evaluate the models' cross-domain performance.

To enable a close comparison with the replicated study, we computed the ROC-AUC score~\cite{sofaer2019area} for each of our fine-tuned models per dataset, except for LaMini-Flan-T5-77M and ChatGPT-4o-mini that produce text outputs instead of probability scores, preventing the computation of ROC-AUC values. Furthermore, we also computed metrics such as accuracy, precision, recall, and F1-score, which are the common metrics associated with classification tasks~\cite{goutte2005,naidu2023review}. 

The parameters to all other transformers were kept at their default values, except to ChatGPT-4o-mini, which was set with \textit{temperature} equal to 0 for reproducibility and \textit{max\_tokens} fixed at 10 to constrain the output to short labels. 
We evaluated each model ten times (i.e., 10 independent runs) on each dataset and report the mean values of the considered metrics. These repeated runs were also used to assess statistical significance through Wilcoxon signed-rank tests~\cite{wilcoxon1945}, allowing us to determine whether the observed differences in performance across datasets or augmentation strategies are significant.



Except for ChatGPT-4o-mini, which relies on the OpenAI APIs,\footnote{https://github.com/openai/openai-python} all experiments were conducted on a machine equipped with an Intel Core i9 (13th Generation) processor, an NVIDIA GeForce RTX 4090 GPU, and running Ubuntu 24.04.1 LTS. 
On average, fine-tuning BERT, RoBERTa, and XLNet required approximately 30 minutes per epoch on the combined dataset and about 15 minutes on the \textit{Questions Only} dataset. For LaMini-Flan-T5-77M, fine-tuning on the combined dataset took roughly 2 hours compared to about 1 hour on the \textit{Questions Only} dataset. To prevent model overfitting, we utilized early stopping.

Similar to prior work~\cite{mahadi2022conclusion}, we focused on evaluating the inherent generalization ability of each model architecture under standard configurations rather than maximizing performance through extensive hyperparameter tuning. Using default settings ensures a fair and consistent comparison across models, reduces the risk of dataset-specific overfitting, and improves reproducibility of our results. Moreover, large-scale tuning across multiple models and datasets would introduce substantial computational cost and confound the analysis by attributing improvements to optimization choices rather than architectural differences. Therefore, we adopted the commonly used default hyperparameters provided by the HuggingFace implementations for all models.

\section{Results}
\label{sec:results}
This section describes the results from testing our models on the ``within'' domain (Stack Overflow) as well as ``cross-domain'' datasets from GitHub issues, discussions, and commit messages (see Table~\ref{tab:datasets}).

\subsection{Model Performance Within Dataset}
\label{sec:within_data}

First, we test our initial hypothesis that, when fine-tuned on a combination of questions, comments, and answers, the model's performance is similar to that when fine-tuned on questions only (see Section~\ref{sec:transformers}). Despite the models being fine-tuned on these dataset variants, we evaluated them on the same dataset, which combined questions, comments, and answers, because this allowed us to compare our results with those of~\citet{mahadi2022conclusion}. The combined dataset comprised 200,000 samples, whereas the questions contained only 100,000. We excluded ChatGPT-4o-mini from this evaluation, as it would require an estimated 600,000 API calls (30,000 samples × 2 model variants × 10 runs), which would incur prohibitive computational and monetary costs. Additionally, we noted that fine-tuning ChatGPT-4o-mini on the combined dataset would take approximately 5 hours 30 minutes, whereas fine-tuning on the \textit{Questions Only} dataset would take about 2 hours 30 minutes.


Table~\ref{tab:model_metrics_train_variants} presents the results obtained by fine-tuning the four models on these two variants of data composition. We observe in this table that discarding the answers and comments do not affect the model's performance across any of the metrics. The values for each metric differ by only a small margin. This observation confirms the observed phenomena of model saturation as training set size grows~\cite{zhu2016we}.
In comparison with the results from \citet{mahadi2022conclusion} (i.e the replicated study), our highest ROC-AUC score of 0.958 for RoBERTa on \textit{Combined} and for XLNet on \textit{Questions Only} (see Table~\ref{tab:model_metrics_train_variants}) is greater than their score of 0.881 obtained using a Linear SVM classifier trained on the combined dataset of 200,000 samples. The high values exhibited by the models in our study, particularly precision and recall, imply that the transformer-based models trained on a smaller dataset outperform the traditional ML classifiers in detecting design discussions from natural language text. Thus, the transformer-based models have a superior ability to differentiate between design and non-design discussions when trained and tested in the same domain.

\begin{table}[!tp]
\caption{Performance metrics (mean values) for different models and fine-tuning data variants}
\centering
\addtolength{\tabcolsep}{-1pt}
\begin{tabular}{l|l|c|c|c|c}
\toprule
\textbf{Model} & \textbf{Train Variant} & \textbf{AC} & \textbf{ROC-AUC} & \textbf{R} & \textbf{P}\\
\midrule
\multirow{2}{*}{XLNet} 
  & Combined        & 0.898 & 0.955 & 0.916 & 0.885 \\
  & Questions Only  & 0.899 & 0.958 & 0.879 & 0.915 \\
\midrule
\multirow{2}{*}{RoBERTa} 
  & Combined        & 0.898 & 0.958 & 0.897 & 0.899 \\
  & Questions Only  & 0.887 & 0.947 & 0.912 & 0.868 \\
\midrule
\multirow{2}{*}{BERT} 
  & Combined        & 0.900 & 0.954 & 0.899 & 0.902 \\
  & Questions Only  & 0.893 & 0.946 & 0.881 & 0.903 \\
\midrule
\multirow{2}{*}{LaMini} 
  & Combined        & 0.894 & - & 0.894 & 0.894 \\
  & Questions Only  & 0.886 & - & 0.905 & 0.885 \\
\bottomrule
\end{tabular}

AC: Accuracy; R: Recall; P: Precision
\label{tab:model_metrics_train_variants}
\end{table}

To assess whether fine-tuning on \textit{Combined} data versus \textit{Questions Only} data significantly impacts model performance, we applied the Wilcoxon signed-rank test~\cite{wilcoxon1945} across multiple evaluation metrics.  The results revealed statistically significant differences in all pair comparisons (\textit{Combined} vs. \textit{Questions Only}) for all transformers and metrics, except for LaMini-Flan-T5's precision and XLNet's accuracy. Despite this statistical significance, the mean difference between \textit{Combined} and \textit{Questions Only} trained models ranged from –0.0110 to 0.0004 for accuracy, –0.0306 to 0.0303 for precision, –0.0368 to 0.01549 for recall, and –0.0101 to 0.0030 for ROC-AUC. We conclude that discarding comments and answers results in halving model training time, as noted previously in the model training time differences. The results for each metric are provided in full tables in the supplementary materials~\cite{supplementary}.

\begin{tcolorbox}
\textbf{Summary of results Within Dataset:}
The results confirm our hypothesis that discarding answers and comments from the fine-tuning data does not significantly affect model performance. This suggests that question text alone is sufficient for reliably detecting design discussions, offering a more streamlined and efficient approach to data preparation, while reducing fine-tuning time.
Our results also reveal that transformer-based models perform better than traditional classifiers used in prior work. in detecting design-related discussions when both training and testing data originate from the same domain, with ROC-AUC scores as high as 0.958. This makes them suitable replacement for traditional classifiers from the same domain.
\end{tcolorbox}

\subsection{Model Performance Cross-Domain Dataset}
\label{sec:cross_domain}

Having observed satisfactory model performance with the Stack Overflow dataset, we proceed to evaluate the fine-tuned model's performance on datasets from different domains (cross-domain).

\begin{figure}[!tp]
    \centering
    \includegraphics[width=.7\linewidth]{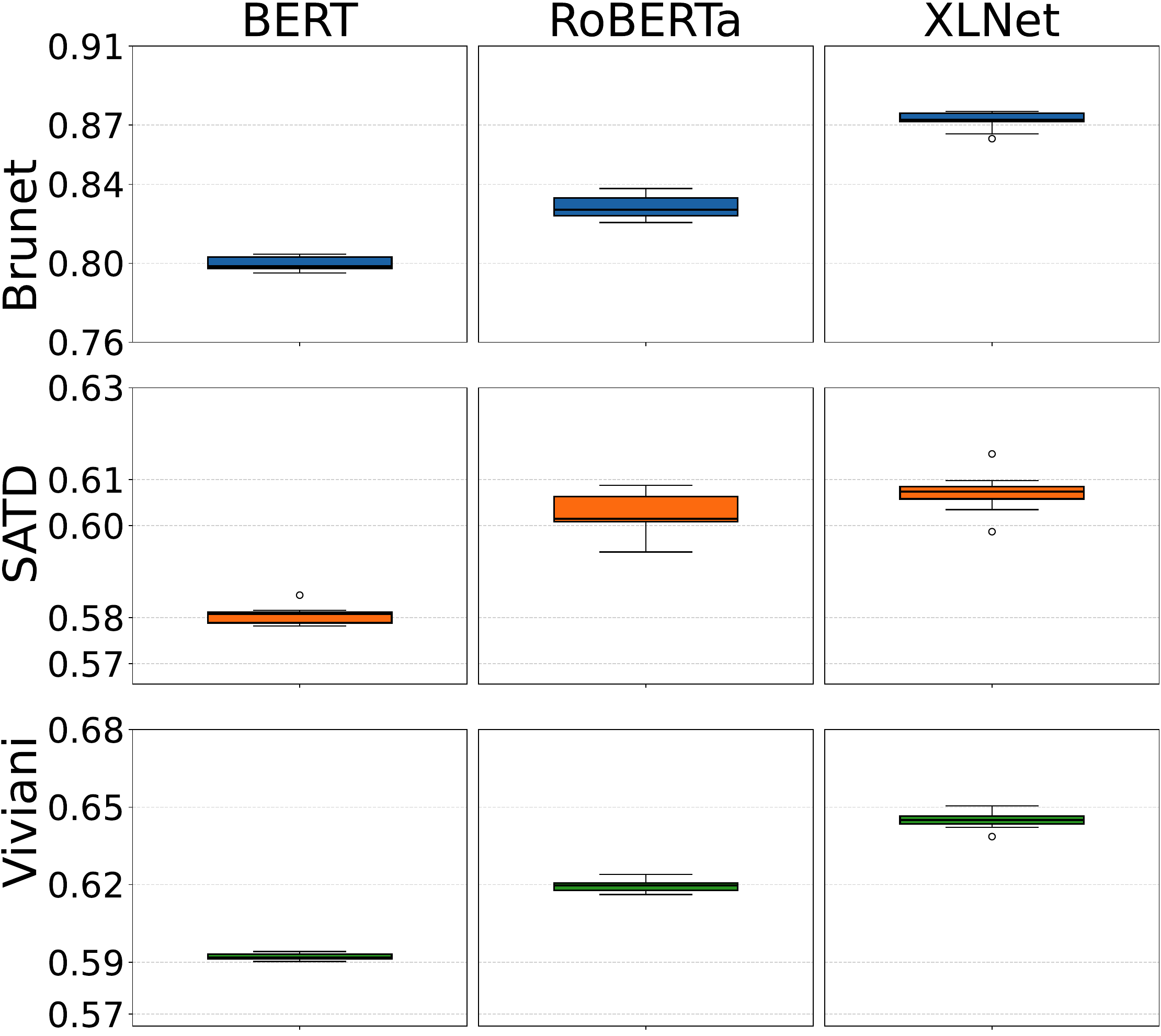}
    \caption{Boxplots of ROC-AUC Scores for the 10 independent runs per model (columns) and per dataset (rows).}
    \label{fig:roc_auc_scores}
\end{figure}

Figure~\ref{fig:roc_auc_scores} shows the ROC-AUC scores recorded when our models fine-tuned on \textit{Questions Only} are applied to the cross-domain datasets for design discussion classification. For cross-domain ROC-AUC score, \citet{mahadi2022conclusion} recorded their best scores for the dataset provided by Brunet~\cite{brunet2014developers}, Viviani~\cite{viviani2019locating}, and SATD~\cite{da2017using} to be 0.632, 0.513, and  0.505, respectively. However, for these same datasets, the models used in our study achieved mean scores of 0.872 (Brunet: XLNet questions only), 0.645 (Viviani: XLNet questions only), and 0.607 (SATD: RoBERTa).
It is also worth mentioning that the models we fine-tuned on the \textit{Questions Only} dataset perform closely to their counterparts fine-tuned on the \textit{Combined} dataset for these cross-domain datasets (see Section~\ref{sec:within_data}). This suggests that the model fine-tuned on less information, namely the questions without comments, gives an equivalent level of understanding and performance when classifying design discussions of texts from different domains. 
Although we do not present the results for the \textit{Combined} dataset in the paper, the details are available in our supplementary materials~\cite{supplementary}.

Another noteworthy consideration is that in the prior work by \citet{mahadi2022conclusion}, the authors also tried to improve models' performance by using data augmentation. Data augmentation aimed to transfer context from the training data to the validation data. 
After this improvement, the Linear SVM classifier reached the best ROC-AUC of 0.798, Precision of 0.647, and Recall of 0.901 for the dataset provided by \citet{brunet2014developers}. Interestingly, our XLNet model fine-tuned on the \textit{Questions Only} dataset, without any hyperparameter tuning or data augmentation, obtained a greater ROC-AUC score, namely 0.872, on the same dataset. 
This can be observed in Figure~\ref{fig:roc_auc_scores}, for  ROC-AUC.
Table~\ref{tab:cross_domain_no_inject} reports the precision and recall scores, which are 0.665 and 0.679, respectively. Although our model achieves higher ROC-AUC and precision compared to \citet{mahadi2022conclusion}, its recall is comparatively lower. Notably, XLNet demonstrated strong performance in identifying design discussions, closely aligning with the manual annotations in the cross-domain dataset provided by~\citet{brunet2014developers}, despite not being trained on data from the same domain (i.e., GitHub pull request discussions). We attribute the lower recall to the model's limited ability to fully capture the nuanced context under which the original annotators classified discussions as design-related, leading to the model failing to identify some design-related discussions.

Our transformer-based classifiers yielded lower performance on the Viviani and SATD datasets, as shown in Table~\ref{tab:cross_domain_no_inject}. 
This outcome indicates limited generalization across datasets originating from different discussion domains, such as GitHub, despite the presence of valuable design discussions. 
We attribute this to variations in dataset annotation guidelines. 
For instance, the Viviani dataset defines design discussions through ``the use of speculative language or the presence of rationale supporting statements''~\cite{viviani2019locating}. 
In contrast, the SATD dataset identifies design-related content through concepts such as ``misplaced code'', ``lack of abstraction'', ``long methods'', ``poor implementation'', and ``temporary solutions'' \cite{da2017using}. Also, Brunet et al.~\cite{brunet2014developers} focus on textual elements associated with software structural design.

Although transformer models outperform traditional ML classifiers on cross-domain data, their overall performance, particularly in terms of balanced precision and recall, declines noticeably in comparison to within dataset. For example, our RoBERTa model achieves a precision of 0.454 and a recall of 0.790 when evaluated on the Brunet dataset, meaning that while the model captures a large proportion of relevant instances (high recall), it also includes a notable number of false positives (moderate precision). Thus, caution is advised when applying models trained on one platform (e.g., Stack Overflow) to another (e.g., GitHub developer discussion), as domain adaptation or fine-tuning may be necessary.

The results presented above are the basis to answer RQ1. We provide the answer and the associated discussion in Section~\ref{sec:discussion}.

\begin{table}[!tp]
\centering
\addtolength{\tabcolsep}{-1pt}
\caption{Cross-domain performance of models fine-tuned on question-only data.}
\begin{tabular}{l|l|c|c|c|c|c}
\toprule
\textbf{Model} & \textbf{Dataset} & \textbf{AC} & \textbf{ROC-AUC} & \textbf{R} & \textbf{P} & \textbf{F1} \\
\midrule
\multirow{3}{*}{BERT} 
  & Brunet   & 0.697 & 0.800 & 0.767 & 0.435 & 0.555 \\
  & SATD     & 0.533 & 0.581 & 0.483 & 0.748 & 0.587 \\
  & Viviani  & 0.578 & 0.592 & 0.543 & 0.335 & 0.414 \\
\midrule
\multirow{3}{*}{RoBERTa} 
  & Brunet   & 0.715 & 0.828 & 0.790 & 0.454 & 0.577 \\
  & SATD     & 0.526 & 0.602 & 0.456 & 0.758 & 0.569 \\
  & Viviani  & 0.593 & 0.619 & 0.566 & 0.351 & 0.433 \\
\midrule
\multirow{3}{*}{XLNet} 
  & Brunet   & 0.837 & 0.872 & 0.679 & 0.665 & 0.672 \\
  & SATD     & 0.462 & 0.607 & 0.293 & 0.793 & 0.428 \\
  & Viviani  & 0.682 & 0.645 & 0.348 & 0.408 & 0.375 \\
\midrule
\multirow{3}{*}{ChatGPT} 
  & Brunet   & 0.731 & --     & 0.684 & 0.467 & 0.555 \\
  & SATD     & 0.562 & --     & 0.543 & 0.750 & 0.630 \\
  & Viviani  & 0.521 & --     & 0.610 & 0.310 & 0.411 \\
\midrule
\multirow{3}{*}{\makecell{LaMini\\Flan-T5}} 
  & Brunet   & 0.807 & --     & 0.536 & 0.625 & 0.577 \\
  & SATD     & 0.444 & --     & 0.276 & 0.765 & 0.406 \\
  & Viviani  & 0.687 & --     & 0.280 & 0.400 & 0.329 \\
\bottomrule
\end{tabular}
\label{tab:cross_domain_no_inject}
\end{table}

\subsection{Model Performance With Similar Words Injection}
\label{sec:similar_word_injection}
\begin{figure}
    \centering
    \includegraphics[width=0.7\linewidth]{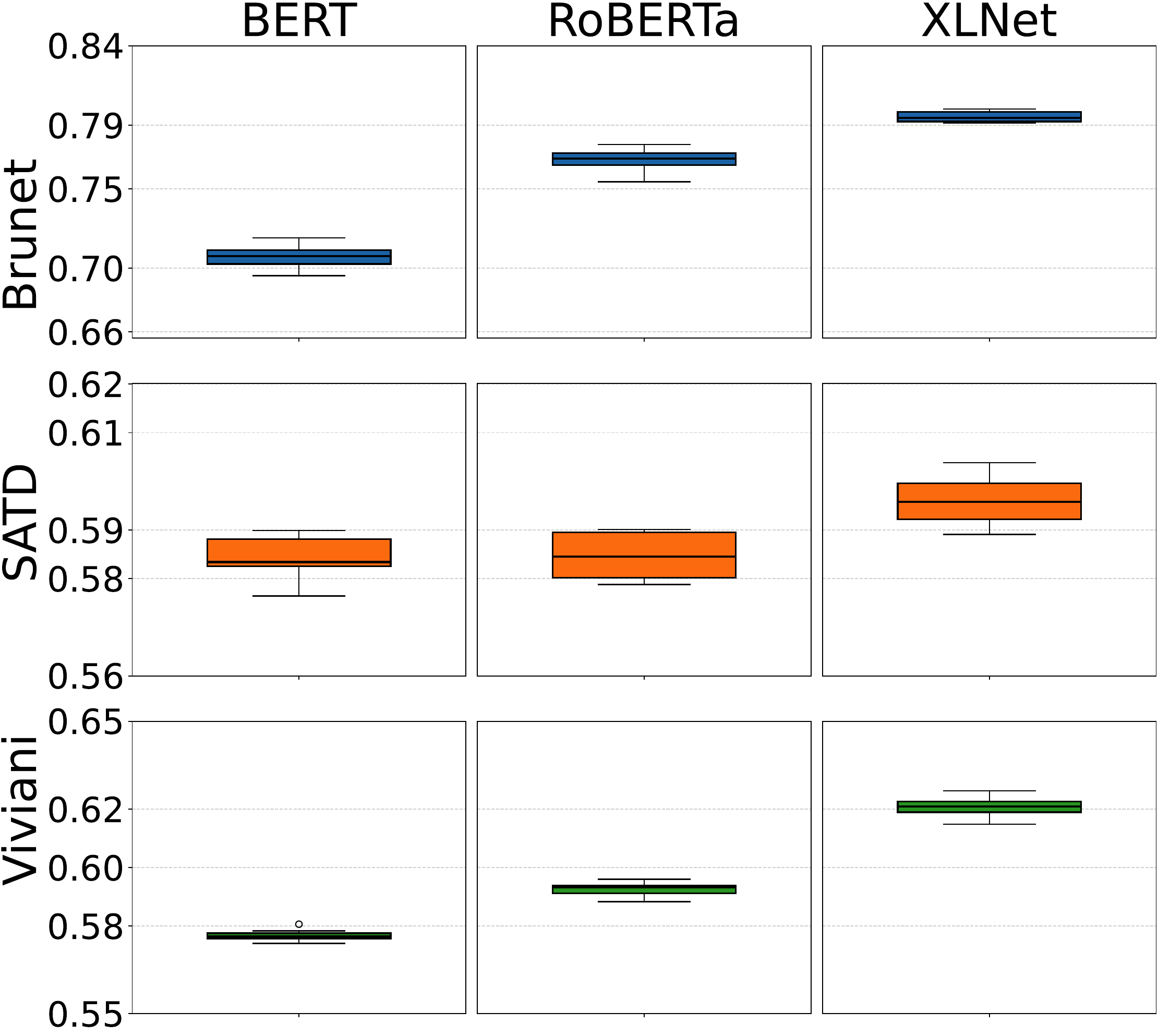}
    \caption{Boxplots of ROC-AUC Scores for the 10 indepent runs per model (columns) and per dataset (rows) with similar words injected.}
    \label{fig:roc_auc_scores_simlar_words_injected}
\end{figure}

We experimented with transferring contextual information from the fine-tuning data to the test data to investigate whether transformer-based classifiers could perform better on cross-domain datasets using word synonym injection. Our models were fine-tuned exclusively on Stack Overflow data. To this end, we extracted words from the fine-tuning set, filtered out stop words (as they typically do not carry meaningful semantic content), and injected the remaining terms into each cross-domain dataset as synonyms following prior data augmentation work~\cite{wei2019eda}, using a newly initialized pre-trained XLNet model. Unlike traditional synonym replacement approaches based on cosine similarity between static word embeddings, which identify globally similar words independent of context, our method relies on contextualized masked language modeling. XLNet predicts substitute tokens conditioned on the full sentence, allowing injected words to better preserve the original semantic intent while aligning lexical usage with the training domain. The XLNet model was not fine-tuned on our datasets and was used solely as a fixed augmentation mechanism to prevent information leakage.

The results are summarized in Table~\ref{tab:injected_similar_words}. Figure~\ref{fig:roc_auc_scores_simlar_words_injected} presents the box plots for the ROC-AUC scores. When comparing the ROC-AUC scores in Tables~\ref{tab:cross_domain_no_inject} and~\ref{tab:injected_similar_words} we can observe that the values for XLNet on the Brunet dataset varied from 0.872 to 0.795, in SATD from 0.607 to 0.596, and in Viviani from 0.645 to 0.621. We can corroborate this observation with analysis of boxplots in Figures~\ref{fig:roc_auc_scores} and~\ref{fig:roc_auc_scores_simlar_words_injected}. We attribute negligible variation to the models' strong ability to generalize based on the semantic content of input sentences. Since the models have already learned robust representations of sentence meaning and semantics during fine-tuning, replacing certain words with their semantically similar counterparts did not significantly alter the underlying sentence representations. Consequently, the predictions remained largely unaffected, indicating that the models were not too sensitive to surface-level lexical variations. Similar findings have been reported with pretrained models like BERT and RoBERTa~\cite{longpre2020effective}. Additionaly, our findings corroborate the observations in prior work, which found that the performance of classifiers decrease or show only marginal increases after similar word injection~\cite{mahadi2022conclusion}.

To draw robust conclusions about the impact of synonym-based augmentation, we applied Wilcoxon signed-rank tests across various metrics. While the tests revealed statistically significant differences (p-values lower than 0.05), these did not translate into meaningful performance improvements across the metrics. We found the mean delta for the ROC-AUC metric ranged from –0.0924 to 0.0039, for recall from –0.0988 to 0.1598, for precision from –0.1660 to 0.0128, and for accuracy from –0.0932 to 0.0411. Despite statistically significant differences between similar-word injection and raw data, the SATD dataset, as one example, showed mean accuracy differences of 0.0279 for RoBERTa, –0.0104 for BERT, 0.0347 for XLNet, 0.0158 for ChatGPT, and –0.0650 for LaMini-Flan-T5.

The takeaway is that simple augmentation strategies introduce extra cost without practical benefit, highlighting the need for more advanced approaches. 
An overall summary of the results above are in Section~\ref{sec:discussion}, with the answer to RQ2 and an associated discussion.

\begin{table}[!tp]
\centering
\caption{Performance of models on datasets (questions only) injected with similar words.}
\addtolength{\tabcolsep}{-1pt}
\begin{tabular}{l|l|c|c|c|c|c}
\toprule
\textbf{Model} & \textbf{Dataset} & \textbf{AC} & \textbf{ROC-AUC} & \textbf{R} & \textbf{P} & \textbf{F1} \\
\midrule

\multirow{3}{*}{BERT} 
  & Brunet   & 0.652 & 0.707 & 0.676 & 0.383 & 0.489 \\
  & SATD     & 0.522 & 0.584 & 0.459 & 0.748 & 0.569 \\
  & Viviani  & 0.598 & 0.577 & 0.444 & 0.329 & 0.378 \\
\midrule

\multirow{3}{*}{RoBERTa} 
  & Brunet   & 0.622 & 0.768 & 0.804 & 0.375 & 0.511 \\
  & SATD     & 0.554 & 0.585 & 0.548 & 0.736 & 0.628 \\
  & Viviani  & 0.533 & 0.593 & 0.649 & 0.325 & 0.433 \\
\midrule

\multirow{3}{*}{XLNet} 
  & Brunet   & 0.753 & 0.795 & 0.681 & 0.499 & 0.576 \\
  & SATD     & 0.496 & 0.596 & 0.384 & 0.767 & 0.512 \\
  & Viviani  & 0.646 & 0.621 & 0.426 & 0.373 & 0.397 \\
\midrule

\multirow{3}{*}{ChatGPT} 
  & Brunet   & 0.683 & --     & 0.844 & 0.427 & 0.567 \\
  & SATD     & 0.577 & --     & 0.572 & 0.754 & 0.650 \\
  & Viviani  & 0.531 & --     & 0.648 & 0.323 & 0.431 \\
\midrule

\multirow{3}{*}{\makecell{LaMini\\Flan-T5}} 
  & Brunet   & 0.742 & --     & 0.742 & 0.767 & 0.751 \\
  & SATD     & 0.485 & --     & 0.485 & 0.625 & 0.491 \\
  & Viviani  & 0.648 & --     & 0.648 & 0.650 & 0.649 \\
\bottomrule
\end{tabular}
\label{tab:injected_similar_words}
\end{table}

\subsection{Evaluation with Generative Language Models}
\label{sec:lamini}
\begin{figure}
    \centering
    \includegraphics[width=0.9\textwidth]{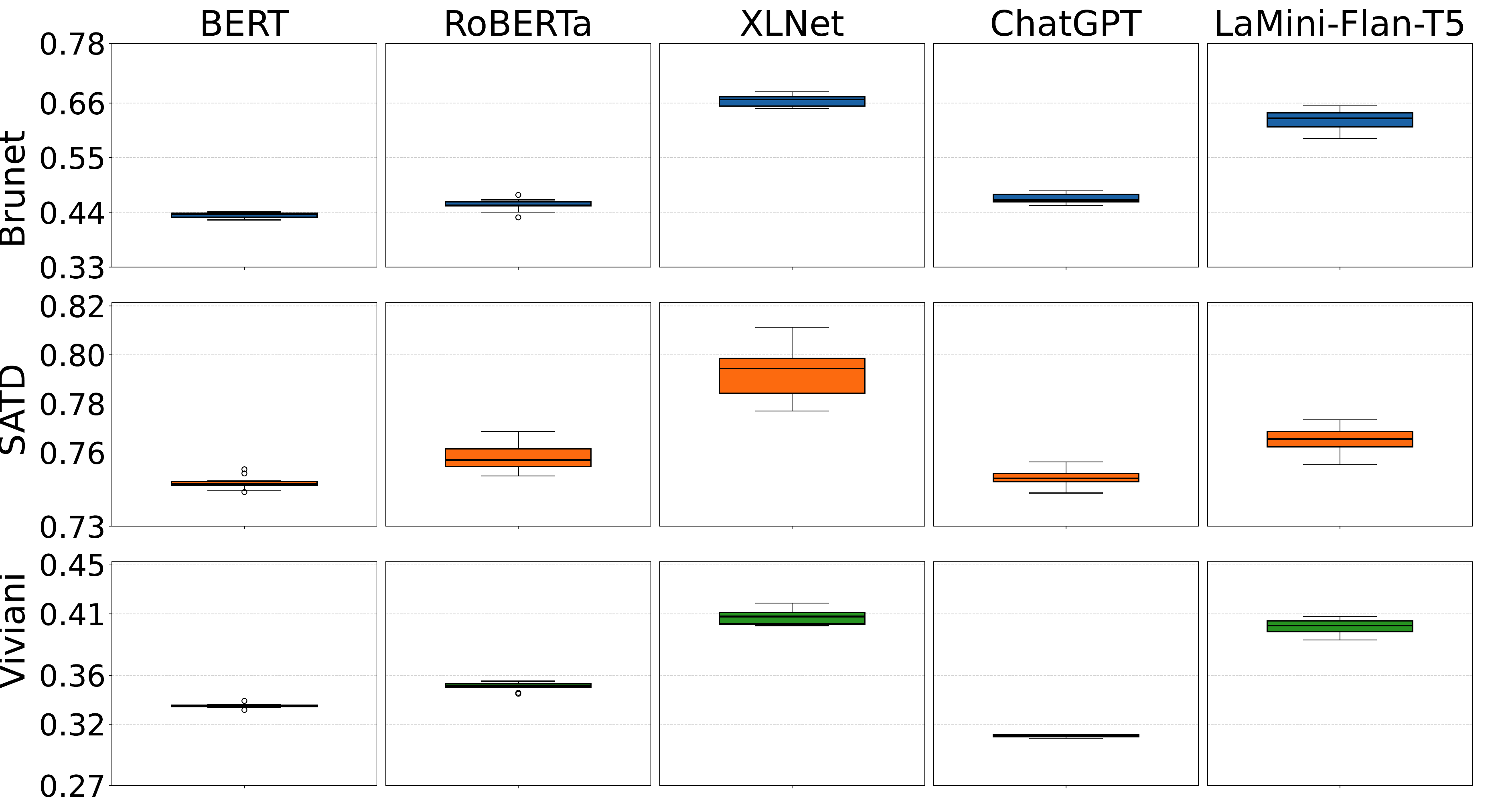}
    \caption{Precision Scores per model per dataset }
    \label{fig:precision_scores_with_flan_t5}
\end{figure}

\begin{figure}
    \centering
    \includegraphics[width=0.9\textwidth]{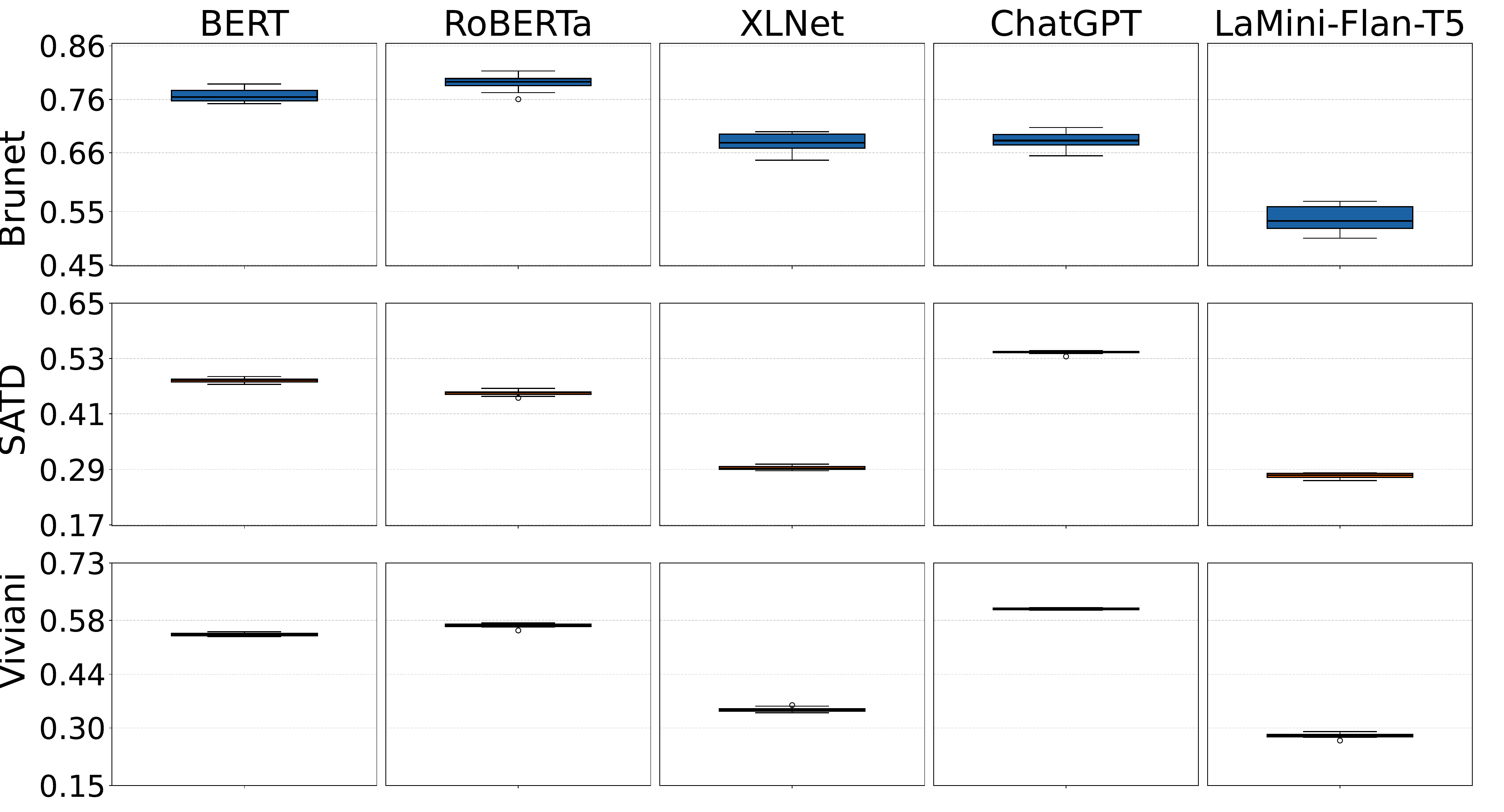}
    \caption{Recall Scores per model per dataset}
    \label{fig:recall_scores_with_flan_t5}
\end{figure}

We further explored the capabilities of state of the art small and large language models for cross-domain text classification. Specifically LaMini-Flan-T5-77M and ChatGPT-4o-mini.
Unlike the other models, LaMini-Flan-T5-77M and ChatGPT-4o-mini output text rather than probability logits, making ROC-AUC not ideal; we therefore report only precision and recall (Tables~\ref{tab:cross_domain_no_inject} and~\ref{tab:injected_similar_words}).
Figures~\ref{fig:precision_scores_with_flan_t5} and~\ref{fig:recall_scores_with_flan_t5} illustrate the precision and recall of the generative language models compared to the other baselines. Since these datasets are imbalanced (Table~\ref{tab:datasets}), we focus on accuracy, precision, and recall and report F1 scores with caution due to the data imbalance. On the Brunet dataset, XLNet achieved the strongest performance with the highest accuracy and precision, while LaMini-Flan-T5-77M performed the weakest due to low recall. For the SATD dataset, ChatGPT-4o-mini attained the most balanced performance, with the highest recall and competitive accuracy, whereas LaMini-Flan-T5-77M again underperformed with very low recall despite relatively high precision. On the Viviani dataset, LaMini-Flan-T5-77M and XLNet were the strongest, with LaMini-Flan-T5-77M obtaining the highest accuracy and XLNet showing competitive recall, while ChatGPT-4o-mini achieved the highest recall but the lowest precision.

\begin{table*}[!tp]
\centering
\caption{Sampled disagreements across datasets (per-model predictions with tick/cross).}
\label{tab:sampled-disagreements}
\addtolength{\tabcolsep}{-4pt}
\begin{tabular}{
    l|   
    r|   
    p{40mm}|   
    c|   
    c|   
    c|   
    c|   
    c|   
    c   
}
\toprule
\textbf{Dataset} & \textbf{ID} & \textbf{Text} & \textbf{\makecell{Orig. \\ Label}} & \textbf{BERT} & \textbf{RoBERTa} & \textbf{XLNet} & \textbf{ChatGPT} & \textbf{\makecell{LaMini \\ FlanT5}} \\
\midrule
SATD & 1585 & is it possible to use more than one variable & general & \cmark & \cmark & \xmark & \xmark & \xmark \\
\midrule
SATD & 3284 & arguments compilers always create irubyobject but we want to use rubyarray concat here as a result this is not efficient since it creates and then later unwraps an array & design & \cmark & \cmark & \xmark & \xmark & \cmark \\
\midrule
BRUNET & 131 & I still think it would make sense to take and stash handOverData during the entire lifetime of the singleton & design & \cmark & \cmark & \cmark & \xmark & \cmark \\
\midrule
BRUNET & 129 & The return value is only used in one place line 195 where the whole bug fix originated & general & \xmark & \xmark & \xmark & \cmark & \xmark \\
\midrule
VIVIANI & 6429 & it is passed as an argument to module wrapper. Not a good idea for \_stream\_readable.js though as it is supposed to be the same as in readable-stream, so only public APIs. & design & \xmark & \cmark & \cmark & \xmark & \cmark \\
\midrule
VIVIANI & 3399 & Could at least have a class attribute with the messages you want to store to test against it, it's more self-contained. & general & \xmark & \cmark & \xmark & \cmark & \xmark \\
\bottomrule
\end{tabular}
\end{table*}

\sloppy
Table~\ref{tab:sampled-disagreements} presents two representative samples from each dataset alongside their ground-truth labels and the predictions produced by the five models. In the SATD dataset, the first instance is clearly non–design-related, which was correctly identified by XLNet, ChatGPT-4o-mini, and LaMini-Flan-T5-77M. The second instance, however, describes an explicit design decision in the code; this was correctly predicted by BERT and RoBERTa, but missed by XLNet and ChatGPT-4o-mini. Turning to the Brunet dataset, the first instance reflects a design decision, yet ChatGPT-4o-mini misclassified it. Similarly, in the second instance, text concerning the usage of a return value tied to a bug fix, ChatGPT-4o-mini again produced an incorrect prediction. For the Viviani dataset, the first instance discusses argument passing within a module wrapper and its suitability for another JavaScript file. This reflects a structural decision and was correctly identified by RoBERTa and XLNet. The second instance, however, highlights a suggestion that a class should maintain an attribute to facilitate self-contained testing. Although this instance is labeled as general, we interpret it as more indicative of a design-related suggestion, aligning with the classifications given by RoBERTa and ChatGPT-4o-mini.

\section{Discussion}
\label{sec:discussion}

This section presents an interpretation of the results, answering the two RQs of our study. Additionally, we describe the implications of our work to software engineering practice and research. Finally, the threats to validity are discussed.

\subsection{Answering RQ1}
\label{sec:rq1}

The results presented in the previous section highlight complementary models with distinct strengths. ChatGPT-4o-mini consistently achieves higher recall across datasets, capturing a greater proportion of design-related discussions than traditional transformer baselines. However, this advantage often comes at the expense of precision, as ChatGPT-4o-mini tends to overpredict design-related labels, occasionally misclassifying general discussions. By contrast, LaMini-Flan-T5-77M, despite its lightweight architecture and smaller parameter footprint, delivers competitive accuracy and strong precision, though its recall is less consistent, particularly in cross-domain settings. Thus, ChatGPT-4o-mini is better suited for contexts where maximizing recall is critical, like exploratory analysis or comprehensive design discussions. LaMini-Flan-T5-77M is a more computationally efficient option with competitive precision, making it ideal for integration into resource-constrained tools such as development assistants or code review pipelines.
From this observation, we present the answer to the RQ1 of our study:

\begin{tcolorbox}
\textbf{RQ1 Answer:}
Transformer-based models varied in effectiveness for detecting design-related discussions. LaMini-Flan-T5-77M and XLNet favored precision, while BERT and RoBERTa leaned toward recall. ChatGPT-4o-mini achieved the highest recall but with reduced precision. These results highlight clear precision–recall trade-offs, making model choice dependent on whether minimizing false positives or maximizing coverage is prioritized.
\end{tcolorbox}

Implications of the findings for RQ1:

\begin{itemize}
    \item In agile workflows with short cycles and limited documentation, ChatGPT-4o-mini can serve as a safety net to prevent design-related discussions from being missed. For example, comments that seem implementation-level (e.g., adding a class attribute for self-containment) often reflect deeper architectural concerns such as encapsulation and maintainability. Capturing these discussions reduces knowledge vaporization and enhances team alignment.

    \item In resource-constrained environments, such as IDE assistants, code review tools, or issue triage systems, LaMini-Flan-T5-77M offers a lightweight yet precise mechanism for surfacing design decisions without overwhelming developers with false positives. This supports more targeted documentation, onboarding, and long-term maintenance.

    \item The variation across models indicates that no single model is universally optimal. Organizations must explicitly decide whether avoiding false positives (precision) or capturing as many relevant discussions as possible (recall) better aligns with their development and compliance needs.
\end{itemize}

\subsection{Answering RQ2}
\label{sec:rq2}

The results of our study (in Section~\ref{sec:similar_word_injection}) show that using synonym-based word injection to enhance cross-domain performance resulted in minimal improvement in our model's performance. For instance, using XLNet on the Brunet dataset, we achieved a ROC-AUC score of 0.872, with a precision of 0.665 and a recall of 0.679. After applying a similar word injection stragegy, the ROC-AUC dropped to 0.795, with precision decreasing to 0.499 and recall slightly increasing to 0.681. Although the statistical tests are significant, the mean differences across metrics are very small, suggesting that this augmentation strategy may not yield meaningful improvements in design discussion detection. Further research is needed to fully understand its potential. The answer to RQ2 is:

\begin{tcolorbox}
\textbf{R2 Answer:}
Similar-word injection led to statistically significant changes in most metrics; however, the actual performance improvements were minimal. The technique did not meaningfully enhance cross-domain detection of design discussions, suggesting that synonym-based data augmentation is not a worthwhile strategy for this task. The limited gains do not justify the additional complexity introduced by the augmentation process. 
\end{tcolorbox}

Implications of the findings for RQ2:

\begin{itemize}
    \item Synonym injection does not provide sufficient benefit for practitioners seeking to improve cross-domain detection of design discussions. The complexity it introduces outweighs the marginal gains.

    \item Future practice should explore more promising methods, such as few-shot learning, contrastive training, or domain-specific augmentation, which may better generalize across heterogeneous datasets.

    \item Since synonym-based augmentation produced statistically significant but practically negligible improvements, relying on such techniques may create a false sense of progress. Practitioners should be cautious when adopting lightweight augmentation strategies without evaluating their real-world impact.
\end{itemize}

\subsection{Threats To Validity}
\label{sec:threats}

We discuss the potential threats to the validity of our study following the framework by \citet{Wohlin2012}. In particular, we elaborate on the threats to construct, internal, external and conclusion validity, and our strategies to mitigate them.

\textbf{Construct validity} concerns whether our measurements and treatments accurately represent the theoretical constructs they are intended to capture. The most significant threat in this category stems from the subjectivity in labeling what constitutes a ``design discussion.'' The definition of design discussion is inherently subjective and may vary depending on the annotator's understanding of the concept, their experience level, and the specific domain or context of the project. This subjectivity may introduce label inconsistencies in the dataset, potentially affecting the model's ability to perform effectively on new or unseen data.
To mitigate this threat, we reused established datasets from prior peer-reviewed studies rather than creating our own annotations, and we manually examined representative samples from each dataset to understand their labeling criteria and identify potential inconsistencies. Additionally, we face a mono-method bias as we relied exclusively on transformer-based approaches for cross-domain classification, which may limit our understanding of the problem space. We partially addressed this by evaluating five different transformer architectures with varying design principles and training strategies, providing a broader perspective within the transformer paradigm.

\textbf{Internal validity} addresses whether a causal relationship can be established between our treatments and outcomes without confounding factors. While we do not seek to establish causal relationships, the model chosen to inject semantically similar words into the cross-domain dataset may itself introduce systematic bias. If the similarity model does not adequately reflect the contextual or domain-specific meaning of words, the substitutions may fail to preserve the intended semantics, leading to context distortion or semantic drift that could mislead the classifier and negatively affect its performance. To assess this threat, we conducted parallel experiments both with and without similar word injection, allowing us to isolate and quantify its impact on model performance. Furthermore, the instrumentation threat manifests through the pre-processed nature of our training data. The primary data used for model fine-tuning had already undergone preprocessing before we accessed it, and we did not have control over the cleaning steps that were applied. Some of the words removed during preprocessing may have influenced the model's performance, as certain domain-relevant terms or contextual clues could have been lost, potentially affecting the model's ability to learn nuanced patterns. While we could not fully mitigate this threat due to dataset reuse, we performed minimal additional preprocessing to avoid compounding the issue.

\textbf{External validity} regards the generalizability of our findings beyond the specific context of our study. The primary threat here relates to experimental configuration. The datasets used for cross-domain evaluation (GitHub pull requests, issues, and commit messages) represent only a subset of possible software development communication channels, potentially limiting the broader applicability of our findings to other contexts such as Slack conversations, email threads, or documentation wikis. We note that this inherent limitation, we still evaluated our models across multiple distinct communication channels. This diversity provides evidence that our approach has the potential to transfer across various text formats and discussion styles commonly found in software development.

Lastly, \textbf{conclusion validity} concerns whether the relationships observed in our data are statistically sound. One threat in this category relates to our limited emphasis on hyperparameter tuning. As this study is a conceptual replication, the primary focus was on evaluating the core idea under different conditions and optimizing model performance. However, we did not perform an extensive search over fine-tuning hyperparameters such as learning rate, batch size, or number of epochs. While this aligns with the goals of our study, more refined tuning could potentially impact the observed effects. We mitigated potential overfitting by using early stopping via PyTorch during fine-tuning, and we used consistent default configurations across all models to ensure fair comparisons. Additionally, the reliability of our measures may be affected by the use of pre-trained models with default configurations, which might not be optimal for the specific task of design discussion detection. To address this threat, we employed tests to evaluate the statistical significance of differences, ensuring our conclusions are based on sound statistical analysis rather than numerical observations alone.

\section{Conclusion}
\label{sec:conclusion}

In this study, we conducted a conceptual replication to investigate the effectiveness of transformer-based models—namely BERT, RoBERTa, XLNet, LaMini-Flan-T5-77M, and ChatGPT-4o-mini—in identifying software design discussions across diverse sources of developer communication. By fine-tuning the transformer models on Stack Overflow data and evaluating them on cross-domain sources such as commit messages, GitHub pull requests, and issue discussion threads, we derived several key insights to inform future research and practical applications, namely: (i) domain-specific training matters, (ii) cross-domain performance degrades, (iii) synonym-based data augmentation is largely ineffective and this confirms the findings from prior work, (iv) LaMini-Flan-T5-77M offers a lightweight and precision-oriented option, and (v) ChatGPT-4o-mini consistently achieves high recall, making it effective in maximizing coverage of design-related discussions. These findings demonstrate that large language models, including ChatGPT-4o-mini, can effectively identify software architectural design discussions from natural language text such as commit messages, pull requests, and issue discussions. In future work, we aim to develop tools to help extract meaningful insights from the discussions flagged as design-related, leveraging them to analyze the evolution of legacy systems and to inform the planning of targeted, data-driven software modernization strategies.


\section*{Data Availability}
All source code, collected data, and complementary results are in the supplementary material~\cite{supplementary}.


%
%









\bibliographystyle{spbasic}      
\bibliography{references}

%
%

\end{document}


\definecolor{cellh}{gray}{1.0} 

\newcommand{\la}[1]{{\textcolor{magenta}{~[~\textbf{LA}: \textit{#1} ]}}}
\newcommand{\wk}[1]{{\textcolor{red}{~[~\textbf{WK}: \textit{#1} ]}}}

\title{Assessing the Bug-Proneness of Refactored Code:\\A Longitudinal Multi-Project Study}

\author{\IEEEauthorblockN{Anonymous Author(s)}}

\maketitle
This document presents supplementary material for the paper “Assessing the Bug-Proneness of Refactored Code:\\A Longitudinal Multi-Project Study”.

\begin{table*}[h!]
\centering
\caption{Analyzed Refactoring Types}
\def\arraystretch{2.0}
\begin{tabular}{m{4cm} m{6cm} m{7cm}}
\toprule
\textbf{Refactoring Type} & \textbf{Problem}  & \textbf{Solution}  \\
\midrule
Extract Interface & Several clients use the same subset of a class’s
interface, or two classes have part of their interfaces in common & Extract the subset into an interface\\
\hline
Extract Method & A code fragment can be grouped together & Turn the fragment into a method whose name
explains the purpose of the method \\
\hline
Extract Superclass & There are two classes with similar features & Create a superclass and move the common
features to the superclass \\
\hline
Inline Method &  When a method body is more obvious than the
method & Replace calls to the method with the method’s
content and delete the method itself\\
\hline
Move Field  & A field is, or will be, used by another class more
than the class on which it is defined & Create a new field in the target class, and change
all its users\\
\hline
Move Class  & Your class belongs to a package that other classes
unrelated to it & Move the class to a related package or create a
new package if required for further use\\
\hline
Move Method  &  A method is, or will be, using or used by more
features of another class than the class in which
it is deffined & Create a new method with a similar body in the
class it uses most. Either turn the old method
into a simple delegation, or remove it altogether\\
\hline
Pull Up Field  &  Two subclasses have the same field & Move the field to the superclass \\
\hline
Pull Up Method  &  There are methods with identical results on
subclasses & Move them to the superclass \\
\hline
Push Down Field  &  A field is used only by some subclasses & Move the field to those subclasses\\
\hline
Push Down Method  &  The behavior on a superclass is relevant only for
some of its subclasses & Move it to those subclasses\\
\hline
Rename Class  &  The name of the class does not reveal its purpose & Change the name of the class and update all
callers\\
\hline
Rename Method  &  The name of a method does not reveal its purpose & Change the name of the method and update all
callers\\
\midrule
\end{tabular}
\label{tab:selected-projects}
\end{table*}

\begin{table*}[t!]
\centering
\caption{Analyzed Code Smell Types}
\def\arraystretch{2.0}
\begin{tabular}{m{4cm} m{13cm}}
\toprule
\textbf{Smell Type} & \textbf{Description} \\
        \hline
        Brain Class & Long and complex class that centralizes the intelligence of the system \\
        \hline
        Brain Method & Long and complex method that centralizes the intelligence of a class \\
        \hline
        Class Data Should Be Private & A class exposing its fields, violating the principle of data hiding \\
        \hline
        Complex Class & A class having at least one method with high cyclomatic complexity \\
        \hline
        Data Class & Classes that have only fields and accessor methods \\
        \hline
        Dispersed Coupling & A method that accesses many code elements, and the accessed elements are dispersed among many classes \\
        \hline
        Feature Envy & A method that is more interested in a class other than the one it actually is in \\
        \hline
        God Class & When a class centralizes the system functionality \\
        \hline
        Intensive Coupling & A method that has tight coupling with other methods, and these coupled methods are defined in the context of a few classes \\
        \hline
        Lazy Class & A class with very small dimensions, few methods, and low complexity \\
        \hline
        Long Method & A method that is unduly long in terms of lines of code \\
        \hline
        Long Parameter List & A method having a long list of parameters, some of which are avoidable \\
        \hline
        Message Chain & A long chain of method invocations is performed to implement a class functionality \\
        \hline
        Refused Bequest & A class redefining most of the inherited methods, signaling a wrong hierarchy \\
        \hline
        Shotgun Surgery & When a change demands many small changes across several different classes \\
        \hline
        Spaghetti Code & A class implementing complex methods that interact with each other, without parameters, using global variables \\
        \hline
        Speculative Generality & An abstract class with very few children classes using its methods \\
        \hline
    \end{tabular}
\end{table*}
